\documentclass[twocolumn,superscriptaddress]{revtex4-2}

\usepackage{graphicx}
\usepackage{bm}
\usepackage{xcolor}
\usepackage{amsmath}

\begin{document}


\title{Anisotropic neutron star crust, solar system mountains, and gravitational waves}

\author{J.A. Morales}
\email{jormoral@iu.edu}
\affiliation{ Center for the Exploration of Energy and Matter and Department of Physics, Indiana University, Bloomington, IN 47405, USA }
\affiliation{ Max Planck Institute for Gravitational Physics (Albert Einstein Institute), Callinstrasse 38, 30167, Hannover, Germany }

\author{C.J. Horowitz}
 \email{horowit@indiana.edu}
\affiliation{Facility for Rare Isotope Beams, Michigan State University, East Lansing, Michigan 48824, USA}
\affiliation{ Center for the Exploration of Energy and Matter and Department of Physics, Indiana University, Bloomington, IN 47405, USA }

\date{\today}

\begin{abstract}
``Mountains” or non-axisymmetric deformations of rotating neutron stars (NS) efficiently radiate gravitational waves (GW).  We consider analogies between NS mountains and surface features of solar system bodies.  Both NS and moons such as Europa or Enceladus have thin crusts over deep oceans while Mercury has a thin crust over a large metallic core.  Thin sheets may wrinkle in universal ways.  Europa has linear features, Enceladus has ``Tiger" stripes, and Mercury has lobate scarps.  NS may have analogous features.   The innermost inner core of the Earth is anisotropic with a shear modulus that depends on direction.  If NS crust material is also anisotropic this will produce an ellipticity, when the crust is stressed, that 
grows with spin frequency.  This yields a braking index (log derivative of spin down rate assuming only GW spin down) very different from $n=5$  and could explain the maximum spin observed for neutron stars and a possible minimum ellipticity of millisecond pulsars. 
\end{abstract}


\maketitle


The opening of the gravitational wave (GW) sky is an historic time.  We have observed GW from black hole \cite{PhysRevLett.116.061102} and neutron star \cite{PhysRevLett.119.161101} mergers.   Galileo opened the electromagnetic sky and its extraordinary riches. The GW sky, no doubt, contains additional very exciting signals.  Galileo observed mountains on the Moon.  Ongoing searches for continuous GW from “mountains” (large scale deformations) on rotating neutron stars have not yet detected signals. Targeted searches have focused on known pulsars, with known spin frequency and spin down parameters \cite{Abbott2004,Abbott2005,Abbott2007,Abbott2008,Abbott2010,Abadie2011,Abadie2011_2,Aasi2014,Aasi2015,Aasi2015_2,Abbott2017_2,Abbott2017_5,Abbott2018_2,Abbott2019,Abbott2019_2,Abbott2021Moun,Ashok2021,Rajbhandari2021,Abbott2022_1,Abbott2022_2,Abbott2022_3}. On the other hand, directed searches have focused on locations in the sky that are known or suspected to harbor a neutron star, without prior knowledge of neither the frequency nor any spin-down parameter \cite{Abbott2017_3,Abbott2017_4,Abbott2019_3,Dergachev2019,Ming2019,Piccinni2020,Millhouse2020,Lindblom2020,Zhang2021,Ming2022,Owen2022}. Finally, all-sky searches have focused on searching for unknown sources at unknown locations \cite{Abbott2005_2,Abbott2007_2,Abbott2008_2,Abbott2009,Abadie2012,Aasi2013,Abbott2016_2,Abbott2017_6,Abbott2018_3,Dergachev2020,Dergachev2021,Covas2022,Abbott2022_4,Dergachev2023,Steltner2023}.  These various types of searches are improving.  Next generation GW observatories Cosmic Explorer and Einstein Telescope should extend these searches to hundreds to thousands of times more neutron stars \cite{Reed2021,pagliaro2023continuous}.

Neutron stars, like many solar system bodies, have solid crusts. ``Mountains'', or non-axisymmetric deformations of  the crust, radiate gravitational waves as the star rotates \cite{Ushomirsky2000}.  
The amplitude $h_0$ of gravitational wave radiation from a star a distance $d$ away, rotating with rotational frequency $\Omega$ and moment of inertia $I$ is \cite{Rilesreview},
\begin{equation}
h_0 = \frac{16 \pi^2 G}{c^4} \frac{I}{d}\Omega^2 \epsilon\, .
\end{equation}
Here the important unknown is the shape of the star or ellipticity $\epsilon$ defined as the fractional difference in the star's principle moments of inertia,
\begin{equation}
 \epsilon=(I_1-I_2)/I_3\, ,
 \label{eq.epsilon}
\end{equation}
with $3$ the rotation axis.
Large scale deformations in the crust, or mountains, can give rise to a non-zero $\epsilon$. 


An important first step is to calculate the maximum ellipticity that the crust can support.  This involves simulating the breaking strain of neutron star crust material and then determining the maximum ellipticity the crust material can support against the star's gravity.  Molecular dynamics simulations, including the effects of impurities, defects, and grain boundaries, find that the breaking strain of the crust is large, of order 0.1, because the crust is under great pressure that prevents the formation of voids or fractures \cite{Horowitz2009,10.1111/j.1745-3933.2010.00903.x,Caplan2018}.  

Given this breaking strain, the maximum ellipticity can be calculated using the intuitive formalism of Ushomirsky et al. \cite{Ushomirsky2000}.  They assume the crust can be strained near its breaking strain everywhere and write the maximum deformation the crust can support as a simple integral of the crust breaking stress divided by the local gravitational acceleration.  This yields a maximum ellipticitiy $\epsilon_{max}$ of a few $\times$10$^{-6}$.  Note that Gittins et al., using a simplified external force to deform the star, claim a smaller $\epsilon_{max}$ \cite{Gittins2021}.  However, using Gittins et al. formalism with an improved external force, we find a larger $\epsilon_{max}$  consistent with Ushmirsky et al. \cite{10.1093/mnras/stac3058}. We therefore assume $\epsilon_{max}\approx {\rm few}\times 10^{-6}$.  

This value for $\epsilon_{max}$ can be compared to the smallest observed upper limit, for a rapidly spinning nearby pulsar, $\epsilon_{min}\approx {\rm few} \times 10^{-9}$ \cite{Abbott2022}.   This gives a dynamic range of $\epsilon_{max}/\epsilon_{min}\approx 1000$.  We can detect mountains 1000 times smaller than the maximum crust mountain.  

Unfortunately, we do not know the actual size of neutron star mountains and $\epsilon$ for particular stars.  Electromagnetic observations of surface features are very limited.  For example rotational phase resolved X-ray spectroscopy has mapped the shape of hot spots on some pulsars \cite{Miller_2019}.  However, this thermal information does not directly provide elevations or mass distributions.  Furthermore, mountain building mechanisms may be complex and depend on poorly known material properties.  For example, viscoelastic creep, how a stressed elastic medium relaxes with time, may be important for the lifetime of neutron star mountains \cite{10.1111/j.1745-3933.2010.00903.x}.

For a given postulated mechanism, for example temperature dependent electron capture on accreting stars \cite{Ushomirsky2000}, one can estimate the resulting ellipticity.  However, neutron star crusts may be very rich physical systems that can involve many possible deformation mechanisms.  We consider analogies between neutron star mountains and surface features of solar system bodies for two reasons.  First, solar system observations may suggest particular mountain building mechanisms that could produce interesting $\epsilon$ values for neutron stars and lead to detectable gravitational wave radiation.  Second, the great diversity of solar system bodies suggests that neutron star crusts may also be diverse.  Although the analogy between neutron stars and solar system bodies is incomplete, we have unique observations of solar system surface features.  This provides ``ground truth'' for complex mountain building physics.

We consider a large range of solar system planets and moons starting from very general considerations and proceed to more specialized observations and mountain building mechanisms. Perhaps the most basic observation is that solar system moons are extraordinarily diverse.  This was revealed with the first images of the Galilean moons of Jupiter from the Pioneer and Voyager spacecraft.  ``The satellite surfaces display dramatic differences including extensive active volcanism on Io, complex tectonism on Ganymede and possibly Europa, and flattened remnants of enormous impact features on Callisto" \cite{doi:10.1126/science.204.4396.951}.  Not only are these four moons very different from each other, none is similar to our Moon.  If neutron stars are also diverse, this could be promising for gravitational wave searches.  Indeed, we know of several different classes of neutron stars such as pulsars, millisecond pulsars, or magnetars.  Even if some stars are symmetric with small ellipticities, others may be very different and  have larger ellipticities.                   

Many solar system bodies have large scale asymmetries.  For example the near and far sides of the Moon are very different.  Mars is strikingly asymmetric.  Not only is the low smooth northern hemisphere quite different from the high cratered southern hemisphere but great volcanoes and high plateaus make the East very different from the West \cite{doi:10.1126/science.1101812}.  Iapetus, a moon of Saturn, is extremely asymmetric with a very dark leading hemisphere and an extraordinarily bright trailing hemisphere \cite{Iapetus}.  
These several bodies, with large observed asymmetries, provide at least moral support for there also being asymmetric neutron stars.  


Mimas, another moon of Saturn, has a very large crater Herschel \cite{https://doi.org/10.1029/2021GL093247}.   This single feature, by itself, creates a significant ellipticity.   
Single catastrophic events on neutron stars could likewise produce large asymmetries.   For example, an event that melts a significant fraction of the crust could leave a large ``scar" when the crust re-freezes and produce a non-zero ellipticity.

Leptodermous kosmos is a possible Greek translation of thin-skinned worlds.  
Neutron stars have a thin crust, approximately 1 km thick, over a deep liquid core. There are a number of thin-skinned moons in the solar system.  Both Europa and yet another moon of Saturn Enceladus have thin ice crusts over deep oceans.  These moons have characteristic linear surface features.  Indeed the lines on Enceladus look like ``Tiger stripes".
 Neutron stars, with their thin crusts, may have analogous linear surface features.


Accretion can spin up the equatorial bulge of a neutron star and put the crust under tension while EM or GW radiation can spin down the bulge and put the crust under compression. Thin sheets may wrinkle in universal ways. 
Examples of wrinkling under tension include hanging drapes \cite{PhysRevLett.106.224301}, stretched thin sheets \cite{PhysRevLett.90.074302}, or a water drop on a thin sheet \cite{doi:10.1126/science.1144616}.  Compressional examples include wrinkling from thermal contraction mismatch \cite{LAI20105185} or growth induced wrinkling in leafs and flowers \cite{PhysRevLett.91.086105}.    

The planet Mercury has a thin silicate crust over a large metallic core.  Lobate scarps on Mercury are bow shaped ridges that can be hundreds of kilometers long. These are the most prominent tectonic features on the planet with a few hundred to several thousand meters of vertical relief \cite{https://doi.org/10.1029/JB080i017p02478,10.1130/0091-7613(1998)026<0991:TOLSOM>2.3.CO;2,SOLOMON1977135}. The formation of these features is thought to involve the thermal contraction of the core leading to compressional wrinkling of the thin crust \cite{lobate_scarps}.  
Neutron stars that are significantly spun down may have lobate scarp like wrinkles and these could contribute to a nonzero ellipticity.

Recent observations of seismic waves reverberating through the Earth's center find an anisotropic innermost inner core \cite{Anisotropic2023}.  Here the velocity of shear waves in innermost inner core material is observed to depend on direction.  Don Anderson wrote ``Crystals are anisotropic and tend to be oriented by sedimentation, freezing, recrystallization, deformation, and flow. Therefore we expect the solid portions of the earth to be anisotropic to the propagation of seismic waves and material properties \cite{doi:10.1073/pnas.232565899}."  
 
 We postulate that neutron star crust may also be anisotropic.  This provides a new way to break axial symmetry and generate a non-zero $\epsilon$.  Single crystals are anisotropic.  Neutron star crust is believed to form a body centered cubic (bcc) lattice.  A bcc lattice has a small shear modulus for compressing one axis of the lattice (and expanding the other two axis by half as much so as to conserve the volume).  In addition there is an $\approx$ 8 times larger shear modulus for distorting the square lattice into a rhombus  \cite{horowitz2008molecular}.  Thus a single bcc crystal has a large anisotropy and the velocity of shear waves depends strongly on direction \cite{TSANG1983377}.  


In addition, complex nuclear pasta phases are expected over an approximately 100 m region between the crust and the core \cite{PhysRevLett.50.2066,RevModPhys.89.041002}.  This region can be important because it is the densest part of the crust and may contain half of the crust's mass.  Some pasta shapes, such as Lasagna, are strongly anisotropic \cite{PETHICK19987,PhysRevC.101.055802}. 
 
However, one typically assumes macroscopic regions of a neutron star involve large numbers of micro-crystals (or domains) and each domain has a random orientation.  As a result almost all calculations assume an angle-averaged shear modulus \cite{doi:10.1063/1.4993443}, see also \cite{PhysRevC.101.055802}, where the velocity of shear waves is independent of direction.   

We now consider that the micro-crystals may be partially aligned due to re-crystallization or some other mechanism.  For example, material on an accreting neutron star is constantly being both crystallized as new material is added and melted as material is buried to higher densities and dissolves into the core.  This may create one or more regions, that are not negligibly small compared to the size of the star, where crystals are at least partially aligned.  Each region is assumed to have a single orientation determined for example by the random orientation of a first seed crystal.  Alternatively, pasta may form (partially) aligned with respect to the magnetic field with spaghetti forming along B or B in the plane of lasagna sheets \cite{PhysRevC.101.055802}. 
The shear modulus will be anisotropic by an amount that depends on the amount of alignment of the micro-crystals.

\begin{figure}[htb]
    \centering
    \includegraphics[width=0.48\textwidth]{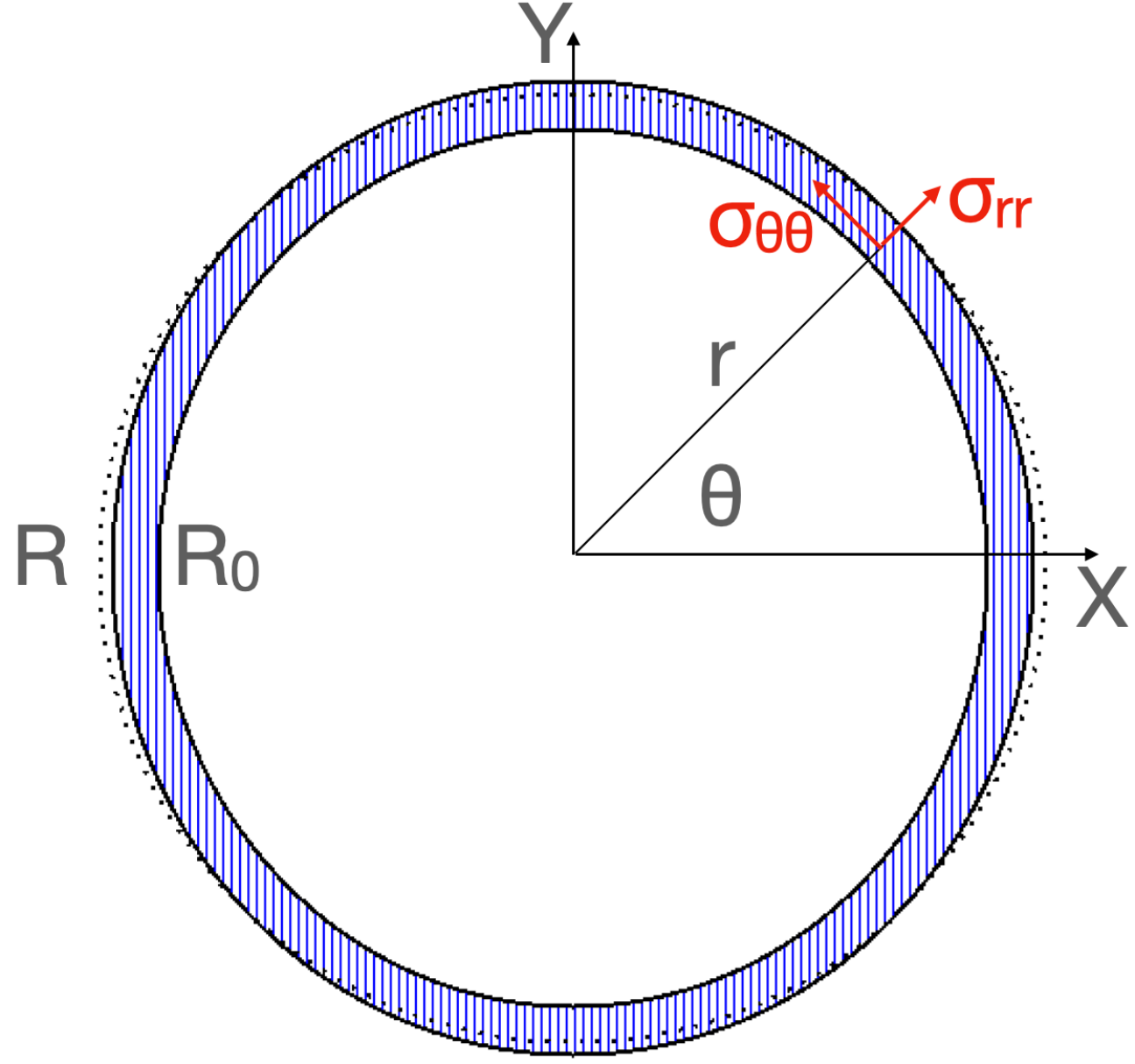}
    \caption{Cut through the equatorial plane of a rotating star.  The crust extends from $R_0$ to $R$ and is slightly anisotropic in the $X$ direction.  
    } \label{fig:diagram}
\end{figure}

We obtain a first estimate of the ellipticity produced by an anisotropic crust with a simple two dimensional calculation.  This order of magnitude result will suffice, given the large uncertainty in the anisotropy of the crust.  We replace a hollow sphere by a hollow cylinder that is assumed thin in the z direction (out of the plane in Fig.~\ref{fig:diagram}).  We consider a thin disc and treat the anisotropy as a first order perturbation to the corresponding axially symmetric constitutive relationships \cite{kelly2013}. A thick disk is expected to yield qualitatively similar results, however the calculation is somewhat more involved.

Figure \ref{fig:diagram} shows the equatorial plane of a rotating neutron star. Elastic perturbations of the stress tensor $\sigma_{ij}$ are related to the strain tensor $\epsilon_{ij}$ by the elasticity of the medium,
\begin{equation}
\sigma_{xx}=\frac{E}{1-\nu^2}(1+\langle\phi\rangle)\epsilon_{xx}+\frac{\nu E}{1-\nu^2}\epsilon_{yy},
\label{eq.sigmaxx}
\end{equation}
\begin{equation}
\sigma_{yy}=\frac{E}{1-\nu^2}\epsilon_{yy}+\frac{\nu E}{1-\nu^2}\epsilon_{xx},
\label{eq.sigmayy}
\end{equation}
and $\sigma_{xy}=E\,\epsilon_{xy}/(1+\nu)$, with $E$ the Young's modulus and $\nu$ the Poisson ratio.  The degree of alignment of micro-crystals in the crust is described by the small parameter $\langle\phi\rangle$.  If $\langle\phi\rangle=0$ the medium is isotropic.  As an example, we consider a partially aligned medium with the symmetries of the Lasagna phase of nuclear pasta \cite{PhysRevC.101.055802}.  The $X$ axis in Fig. \ref{fig:diagram} is normal to the Lasagna planes.  We assume this direction arose from spontaneous symmetry breaking, for example the random oriantation of a seed micro-crystal.   We start with a symmetric medium $\langle\phi\rangle=0$ and then calculate first corrections from $\langle\phi\rangle\ne 0$.  

We note the shear modulus is $\mu=E/[2(1+\nu)]$ and bulk modulus is $K=E/[3(1-2\nu)]$.  We define $\langle \phi\rangle$ to describe the degree of alignment with respect to the shear (or Young's) modulus.  The larger bulk modulus (since $\nu$ is near 0.5) may have a large contribution from the isotropic electron pressure.  This symmetric pressure will tend to reduce (but not eliminate) the ellipticity of the star, see discussion of Eq. \ref{eq.Aresult} below. 

We assume the crust froze while the star was rotating with initial angular frequency $\omega_0$.  If the star is then spun up or spun down to a new angular frequency $\omega$, stresses will be induced according to the equation of motion,
\begin{equation}
\frac{\partial\sigma_{rr}}{\partial r}+\frac{1}{r}(\sigma_{rr}-\sigma_{\theta\theta})=-\rho r(\omega^2-\omega_0^2),
\end{equation}
with $\rho$ the average crust density.  The radial stress is \cite{kelly2013},
\begin{equation}
\sigma_{rr}(r)=\frac{3+\nu}{8}\rho(\omega^2-\omega_0^2)[R^2+R_0^2- r^2-\frac{R^2R_0^2}{r^2}]\, ,
\label{eq.sigmarr}
\end{equation}
and satisfies boundary conditions $\sigma_{rr}(R_0)=\sigma_{rr}(R)=0$ at the inner $R_0$ and outer $R$ radii of the crust.  The angular stress $\sigma_{\theta\theta}$ is \cite{kelly2013},
\begin{equation}
\sigma_{\theta\theta}(r)=\frac{3+\nu}{8}\rho(\omega^2-\omega_0^2)[R^2+R_0^2- \frac{1+3\nu}{3+\nu}r^2+\frac{R^2R_0^2}{r^2}]\, ,
\label{eq.sigmatt}
\end{equation}
and does not vanish at $r=R_0$ or $R$.

We now consider $\langle\phi\rangle\ne 0$.  We rewrite Eqs.~\ref{eq.sigmaxx},\ref{eq.sigmayy} in polar coordinates, invert to obtain strain $\epsilon_{ij}$ as a function of stress, and expand to lowest order in $\langle\phi\rangle$.  We provide a first estimate of the change in strain $\delta\epsilon_{ij}$ with $\langle\phi\rangle$ by using the unperturbed stresses from Eqs~\ref{eq.sigmarr} and \ref{eq.sigmatt} to get,
\begin{equation}
   \begin{bmatrix} 
      \delta\epsilon_{rr} \\
      \delta\epsilon_{\theta\theta}\\
     \end{bmatrix}
   \approx-\frac{\langle\phi\rangle}{E}\begin{bmatrix}\frac{(C^2-S^2\nu)^2}{1-\nu^2} &\frac{C^2S^2(1+\nu)}{1-\nu} 
   \\
 \frac{C^2S^2(1+\nu)}{1-\nu} &\frac{(C^2\nu-S^2)^2}{1-\nu^2} 
 \\
 \end{bmatrix} 
 \begin{bmatrix}
 \sigma_{rr}\\
 \sigma_{\theta\theta}\\
 \end{bmatrix}
 \label{eq.matrix}
  \end{equation}
with $S=\sin\theta$ and $C=\cos\theta$.  We assume a thin crust $R-R_0\ll R$ where $\sigma_{rr}\ll\sigma_{\theta\theta}$ and $\sigma_{\theta\theta}$ is approximately independent of $r$.  This gives $\delta\epsilon_{rr}$ and $\delta\epsilon_{\theta\theta}$ that are also $\approx$ independent of $r$.  The radial displacement is written $u_r=u_r^0+\delta u_r$ where $u_r^0$ is the displacement in the isotropic case $\langle\phi\rangle=0$.  Likewise $u_\theta=u_\theta^0+\delta u_\theta$.  The perturbation $\delta u_r\approx (r-R_0)\delta\epsilon_{rr}(\theta)$ follows from integrating $\partial \delta u_r/\partial r=\delta\epsilon_{rr}$.  The angular displacement follows by integrating $\delta\epsilon_{\theta\theta} =(\partial \delta u_\theta/\partial\theta+ \delta u_r)/r$ to get $\delta u_\theta\approx \int d\theta' [r\delta\epsilon_{\theta\theta}-(r-R_0)\delta\epsilon_{rr}]+C_\theta$ or,
\begin{equation}
\delta u_\theta(\theta)\approx R\int^\theta_0 d\theta'\delta\epsilon_{\theta\theta}+C_\theta\,,
\end{equation} 
assuming $r\gg r-R_0$.  Here the integration constant $C_\theta$ is independent of $\theta$ and does not contribute to moments of inertia.

We now calculate the difference in moments of inertia in Eq.~\ref{eq.epsilon}.  For simplicity we work in two dimensions and actually calculate moments of inertia of a deformed hoop.  This provides an order of magnitude estimate for the ellipticity.  A mass $\rho({\bf r})$ that was initially at ${\bf r}$ is now at ${\bf r}'=(r+u^0_r+\delta u_r)\hat r+(u_\theta^0+\delta u_\theta) \hat\theta+ u_z\hat z$.  The difference in moments of inertia is,
\begin{equation}
I_x-I_y=\int d^3r \rho(r)[({\bf r}'\cdot \hat y)^2-({\bf r}'\cdot\hat x)^2]\,.
\end{equation}
Working to first order in the small displacement $\delta u_\theta(\theta)$, this becomes
$I_x-I_y\approx 4\int d^3r \rho(r)r \delta u_\theta(\theta) \sin\theta\cos\theta$.  We write this as
$I_x-I_y\approx m_{cr}R^2 A$ where $m_{cr}=\int d^3r\rho(r)$ is the mass of the crust and the important angular integral is
\begin{equation}
A=\frac{4}{2\pi}\int_0^{2\pi}d\theta \sin\theta\cos\theta \int_0^\theta d\theta' \delta\epsilon_{\theta\theta}(\theta')\, .
\end{equation}
Finally, dividing by the moment of inertia $I\approx\frac{2}{5}MR^2$, of a star of mass $M$, gives the ellipticity
$\epsilon\approx \frac{5}{2}\frac{m_{cr}}{M}A$. 
Using $\sigma_{\theta\theta}\approx \rho R^2(\omega^2-\omega_0^2)$ from Eq.~\ref{eq.sigmatt}, $\delta\epsilon_{\theta\theta}$ from Eq.~\ref{eq.matrix}, and evaluating the integral gives 
\begin{equation}
\epsilon\approx\frac{5}{16}\Bigl(\frac{5-2\nu+\nu^2}{1-\nu^2}\Bigr)\Bigl(\frac{m_{cr}}{M}\Bigr)\Bigl(\frac{\rho R^2}{E}\Bigr) \langle \phi\rangle (\omega^2-\omega_0^2)\,.
\label{eq.Aresult}\end{equation}

This calculation has been for a thin disk where the {\it stress} is in the xy plane. A thick disk, where the {\it strain} is in the xy plane, gives a similar result.  We expect results for a three dimensional sphere to also be similar.  However, this should be verified in future work.

Equation~\ref{eq.Aresult} involves a ratio of the anisotropic stress perturbation, proportional to $\langle\phi\rangle$, to the isotropic stress perturbation, proportional to the Young's modulus $E$.  In addition, there is a large isotropic stress from the pressure $P$ that is related to the bulk modulus $K$.  To include this, we replace $E$ in Eq. \ref{eq.Aresult} by 3K \footnote{For a nearly spherical system $-r dP/dr=-3V dP/dV=3K$ for a system of volume $V$}.   Finally, we express $3K$ in terms of the centrifugal force that balances gravity.  In hydrostatic equllibrium, $dP/dr=-GM\rho/r^2$.  The gravitational acceleration is balanced by a centrifugal acceleration $GM/R^2\approx \omega_K^2R$ when the star is rotating at the Kepler or break up speed $\omega_K$.  Therefore, we replace $3K=-rdP/dr$ by $\omega_K^2R^2\rho$ to arrive at our main result for the ellipticity of a star with anisotropic material in its crust,
\begin{equation}
\epsilon \approx \frac{m_{cr}}{M} \langle\phi\rangle \frac{\Omega^2-\Omega_0^2}{\Omega_K^2}\, .
\label{eq.epsilonfinal}
\end{equation}
 Note that we have rewritten the angular frequencies $\omega$, $\omega_0$, and $\omega_K$ in terms of rotational frequencies $\Omega=\omega/(2\pi)$ etc.  Here $m_{cr}/M\approx 10^{-2}$ and $\Omega_K\approx 1400$ Hz depending on the equation of state.  We see that $\epsilon$ is a strong function of the rotational frequency $\Omega$ and the initial frequency $\Omega_0$ (when the crust froze).


 
Unfortunately, we do not know the degree of anisotropy of the neutron star crust $\langle\phi\rangle$.  Even a small average can lead to a significant $\epsilon$ and produce observable gravitational waves.  As an example we consider the innermost inner core of the Earth because we have no neutron star observations.   There is a few percent anisotropy in the Earth that extends over the innermost inner core \cite{Anisotropic2023}.  This region has a radius of 300 km and contains about $3\times 10^{-4}$ of the Earth's mass.   If we ignore anisotropies in the rest of the Earth, this corresponds to  $\langle\phi\rangle\approx 10^{-5}$ when averaged over the Earth's total mass.  

Of course this value is not directly relevant for neutron stars.  Nevertheless, if material in a neutron star has an anisotropy of $\langle\phi\rangle\approx 10^{-5}$ when averaged over the crust mass, this would yield $\epsilon\approx 10^{-7}(\Omega^2-\Omega_0^2)/\Omega_K^2$ or $\epsilon\approx 10^{-8}$ for a rapidly rotating object with $(\Omega^2-\Omega_0^2)/\Omega_K^2\approx 0.1$.  Gravitational waves from a nearby star could be detectable for this $\epsilon$ value.

\begin{figure}[htb]
    \centering
    \includegraphics[width=0.44\textwidth]{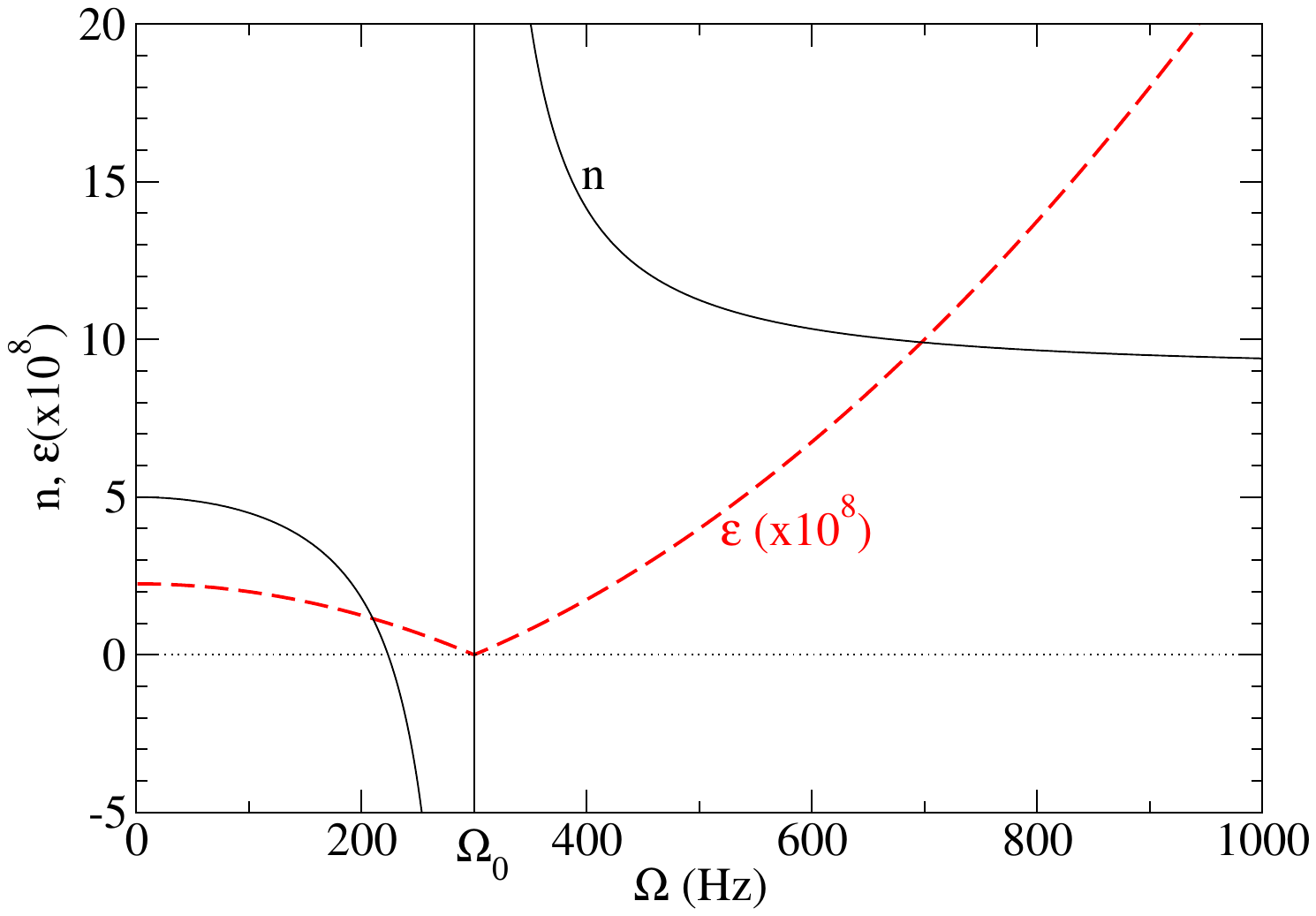}
    \caption{Braking index $n$ (solid black curve) and ellipticity $\epsilon$ (dashed red curve) vs rotational frequency $\Omega$ assuming spin down only from GW radiation and the crust froze while the star was rotating at $\Omega_0=300$ Hz.
    } \label{fig:n}
\end{figure}

The braking index $n=d\ln\dot\omega/d\ln\omega$ describes how the spin down rate $\dot\omega$ depends on rotational frequency.  For simplicity we neglect spin down from electromagnetic radiation.  Spin down from gravitational wave radiation only with a frequency independent $\epsilon$ has $n=5$.  However the strong spin dependence of $\epsilon$ in Eq.~\ref{eq.epsilonfinal} leads to $n=5+4\Omega^2/(\Omega^2-\Omega_0^2)$.  This is very different from 5 as shown in Fig. \ref{fig:n}.  In the limit $\Omega\gg\Omega_0$, $n=9$ and $n$ changes rapidly for $\Omega$ near $\Omega_0$.  Neutron stars that have been spun up since crust formation have $n>9$ while stars that have spun down since crust formation have $n<5$. Finally, $n=0$ at $\Omega=\sqrt{5/9}\,\Omega_0$. For constant $\epsilon$, $|\dot\omega|$ decreases with decreasing $\Omega$.  However as $\Omega$ decreases $\epsilon$ increases (given $\Omega<\Omega_0$) and this increases $|\dot\omega|$.  At $\Omega=\sqrt{5/9}\,\Omega_0$ the two effects cancel and $n=0$.

Torque from gravitational wave radiation could balance the spin up from accretion and limit neutron star spins \cite{Ushomirsky2000}.  Using $\epsilon$ in Eq.~\ref{eq.epsilonfinal}, this torque $N_{gw}$ rises very rapidly  with $\Omega$ so $N_{gw}\propto \Omega^5(\Omega^2-\Omega_0^2)^2\langle\phi\rangle^2$.  If $N_{gw}$ balances the accretion torque $N_a\approx\dot M(GMR)^{1/2}$ then the equilibrium spin,
\begin{equation}
\Omega_{eq}\approx 300\ {\rm Hz}\,(\frac{\dot M}{10^{-8}{\rm M_\odot\, yr}^{-1}})^{1/9}(\frac{10^{-4}}{\langle\phi\rangle})^{2/9}\, ,
\end{equation}
could agree with observed values.  Note that $\Omega_{eq}$ only depends very weakly on the accretion rate $\dot M$ or $\langle\phi\rangle$.   Here we assume $\Omega_{eq}\gg\Omega_0$, $M=1.4M_\odot$, and $R\approx 10$ km.  Because our ellipticity rises strongly with $\Omega$, this torque balance can be achieved with modest $\langle\phi\rangle\approx 10^{-4}$. Furthermore, our mechanism, with a somewhat smaller $\langle\phi\rangle\approx 10^{-5}$, could explain a possible minimum ellipticity $\epsilon\approx 10^{-9}$ suggested by an observed minimum spin down rate for millisecond pulsars \cite{2018ApJ...863L..40W}.

Because the geometry of a thin disc is different from that of a realistic NS, we have performed finite-element simulations of a three-dimensional NS with a polytropic equation of state that confirm the order-of-magnitude estimate of Eq. \ref{eq.epsilonfinal}. The finite-element simulations were performed using \texttt{FEniCSx} software with an element size of order $10^4$ cm for a 1.4 $M_{\odot}$ and 10 km NS \cite{Baratta2023,Scroggs2022,Scroggs2022_2,Alnaes2014}. In addition, the finite-element simulations confirm the proportionality between the ellipticity and $\Omega^2 - \Omega_0^2$ and $\langle \phi \rangle$, as well as the predominance of the contribution from the azimuthal displacement $\delta u_{\theta}$ to the ellipticity. The details of the finite-element simulations will be discussed in a future publication. Our simulations for an $n=1$ polytrope equation of state have an ellipticity
\begin{equation}
\epsilon \approx \frac{m_{cr}}{M} \langle\phi\rangle \frac{\Omega^2-\Omega_0^2}{\Omega_{K,nr}^2} \ ,
\label{eqn:epsilonfinal_nr}
\end{equation}
where $\Omega_{K,nr} \approx 2200 $ Hz is the rotational Kepler frequency for a Newtonian NS with an $n=1$ polytrope equation of state. Note that Eq. \ref{eqn:epsilonfinal_nr} is similar to Eq. \ref{eq.epsilonfinal}.


In conclusion ``mountains” or non-axisymmetric deformations of rotating neutron stars (NS) efficiently radiate gravitational waves (GW).  There are many ongoing searches for continuous GW from such stars.  Present detectors are sensitive, in the best cases, to mountains that are 1000 times smaller than the maximum mountain that the crust can support.  Unfortunately, we do not know the size of NS mountains.  We consider analogies between NS mountains and surface features of solar system bodies.  Here solar system observations can provide ``ground truth" for complex mountain building physics.  Both NS and moons such as Europa or Enceladus have thin crusts over deep oceans while Mercury has a thin crust over a large metallic core.  Thin sheets may wrinkle in universal ways.  Europa has linear features, Enceladus has ``Tiger" stripes, and Mercury has lobate scarps.  NS may have analogous features.   The innermost inner core of the Earth is anisotropic with a shear modulus that depends on direction.  If NS crust material is also anisotropic this could produce a significant ellipticity that 
grows rapidly with increasing rotational frequency.  Gravitational wave emission torques from this ellipticity may limit the spin rate of neutron stars.

\begin{acknowledgments}
Matt Caplan, Cole Miller, Jing Ming, and Ruedi Widmer-Schnidrig are thanked for helpful comments. This work is partially supported by the US Department of Energy grant DE-FG02-87ER40365 and National Science Foundation grant PHY-2116686. 
\end{acknowledgments}





%

\end{document}